# The extinction by dust in the outer parts of spiral galaxies

R. F. Peletier[1,2,3], E.A. Valentijn[4], A.F.M. Moorwood[5], W. Freudling[5,6], J.H. Knapen[3,7], and J.E. Beckman[3]

[1] Isaac Newton Group, Apartado 321, 38700 Santa Cruz de la Palma, Tenerife, Canarias, Spain
[2] Kapteyn Laboratorium, Postbus 800, 9700 AV Groningen, Netherlands
[3] Instituto de Astrofísica de Canarias, 38200 La Laguna, Tenerife, Canarias, Spain
[4] SRON, Postbus 800, 9700 AV Groningen, Netherlands
[5] European Southern Observatory, K. Schwarzschildstr. 2, D-85748 Garching bei München, Germany
[6] Space Telescope - European Coordinating Facility, K. Schwarzschildstr. 2, D-85748 Garching bei München, Germany
[7] Département de Physique, Université de Montréal, C.P. 6128, Succursale Centre Ville, Montréal (Québec), H3C 3J7 Canada

**Abstract.** To investigate the distribution of dust in Sb and Sc galaxies we have analyzed near-infrared and optical surface photometry for an unbiased sample of 37 galaxies. Since light in the $K$-band is very little affected by extinction by dust, the $B - K$ colour is a good indicator of the amount of extinction, and using the colour-inclination relation we can statistically determine the extinction for an average Sb/Sc galaxy.

We find in general a considerable amount of extinction in spiral galaxies in the central regions, all the way out to their effective radii. In the outer parts, at $D_{K,21}$, or at 3 times the typical exponential scale lengths of the stellar distribution, we find a maximum optical depth of 0.5 in $B$ for a face-on galaxy. If we impose the condition that the dust is distributed in the same way as the stars, this upper limit would go down to 0.1.

## 1. Introduction

The analysis of surface photometry in $B$ of large samples has shown that Sb and Sc galaxies suffer from much more extinction than was previously thought (Valentijn 1990). In the centres there is so much extinction that generally the dust optical depth in the visible is larger than unity. There are claims that these spiral galaxies are optically thick even in the outer regions (González-Serrano & Valentijn 1991; Burstein et al. 1991; Bottinelli et al. 1995; Valentijn 1994, hereinafter V94) although these are disputed by several authors (White & Keel 1992; Huizinga & van Albada 1992).

In this *letter* we present the results of a direct way to measure the amount of extinction using radial surface brightness profiles in $B$ and $K$ for a sample of galaxies. Since the extinction in $K$ is generally more than an order of magnitude lower than in $B$, a change in $B - K$, in the absence of a stellar population gradient, indicates only a change in the amount of extinction. Here, we separate the effects of extinction from stellar populations by using the fact that extinction is inclination-dependent, while radial stellar population gradients are not. Thus, by taking a sample of galaxies uniformly distributed in inclination, we can determine statistically the amount of extinction at various places in the galaxy. This cannot be done for optical-optical colours because they suffer from the fact that not only the colour of the underlying stellar population is unknown, but also the extinction law, since dust and stars are mixed, so that the effective extinction law depends on the quantity and the distribution of the dust (see e.g. Evans 1994; Jansen et al. 1994). We present here a study of 37 galaxies, for which the data were presented in Peletier et al. (1994, Paper I). The data allow analysis out to the $21^{st}$ $K$-band isophote, which is close to the $25^{th}$ $B$-band isophote. We analyse in several different ways the results in the centre, at an effective radius (the radius inside which half the light is contained), and at $D_{K,21}$, and discuss the consequences for the optical depth of galaxies at these locations.

## 2. Optical-infrared colour distributions

The sample analyzed here consists of galaxies of morphological type 3-6.4, corresponding to Sb and Sc, with relatively high central Blue surface brightness (20.5-21.5), selected on the basis of the ESO-LV catalogue (Lauberts & Valentijn 1989), i.e. normal late-type spiral galaxies of the types that contain large amounts of extinction (Valentijn 1990). The sample is diameter limited and uniformly distributed in orientation on the sky. A detailed description is given in Paper I.

The advantage of having extinction-free K-band photometry is that it allows us to define a number of radii in each galaxy that depend only on the true stellar distribution, unaffected by extinction. For example, when analyzing the colours at an effective radius in $K$, we do not have to worry how much the measurement of the effective radius itself at various radii is affected by the extinction. We have defined 3 radii at which we study the dependence of $B - K$ as a function of axis ratio. The colours are given in Fig. 1a. On the left we plot the extrapolated disk colour $\mu_{0,B}$ - $\mu_{0,K}$, in the middle the colour as measured at the effective radius in $K$ and, in the right panel, at $\mu_K = 21$. Fig. 1 shows the following effects:

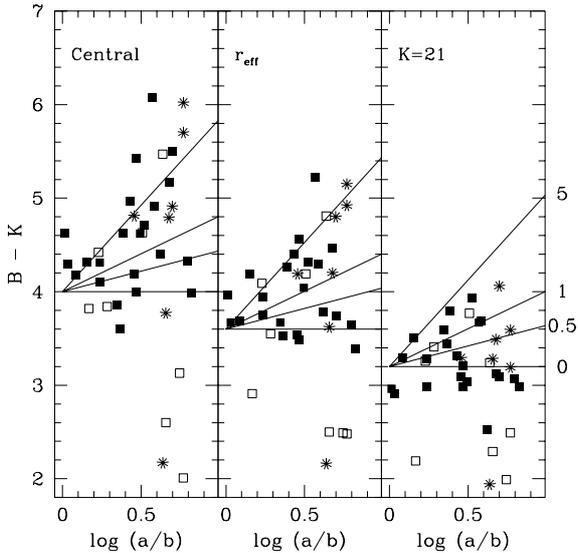

**Fig. 1.** Azimuthally averaged $B - K$ colour at various positions in the galaxy. On the left is shown the extrapolated central disk colour, in the middle the colour at $r_{eff,K}$, and on the right at K=21 mag arcsec$^{-2}$. Galaxies for which no redshift is available, and for which as a result $M_K$ could not be determined, are indicated with asterisks. Filled symbols are given to galaxies brighter than $M_K = -22$, while fainter galaxies are indicated with open symbols. The lines show the colour-dependence as a function of inclination for simple models with central face-on optical depths $\tau^{face}$ of 0, 0.5, 1, and 5.

i) We find that these galaxies are much redder in the centre than in the outer parts. A typical difference in $B - K$ between the centre and $D_{K,21}$ is one magnitude, in some cases more. This is much more than seen in, e.g., elliptical galaxies.

ii) In the centre, the spread in $B - K$ for face-on galaxies is much smaller than for edge-on galaxies. Some edge-ons are much redder than all face-ons, while others are bluer. Fig. 1 shows that the edge-on galaxies that are bluer than all face-on systems are either fainter than $M_K = -22$ or have unknown redshifts (and so could also be fainter than $M_K = -22$). Not all faint systems are edge-on, but the bluest are. We think this is because these galaxies have been misclassified. Edge-on systems are much harder to classify usually than face-ons: small edge-on Sb's on a plate do not look very different from edge-on Sd's. We argue that most of the blue edge-on galaxies here have a type later than 6.5 and have been misclassified in the ESO-LV catalogue (Lauberts & Valentijn 1989). This justifies performing a second selection on the sample, using the $K$-band luminosity $M_K$ in order to obtain a more homogeneous sample for further inclination test studies. (As a matter of fact, one of our conclusions from the K-band photometry is that any sample defined on the basis of its K-band properties is much cleaner than a sample selected on the basis of visual passband photometry. [Paper I].) After having selected $|M_K| > 22$ the situation is much clearer. We find that in the centre (filled dots) and at $r_{eff}$ edge-on galaxies generally get redder strongly with inclination whereas at $D_{K,21}$ no trend of $B - K$ as a function of inclination can be determined. In Fig. 1 we have also displayed the behaviour of a number of models. These are simple models (see Section 3) with a range in face-on extinction. In the central regions, the majority of the galaxies follow the models with a central face-on optical depth of 2 or more. In the outer parts, the transparent models seem to be preferred. One should however note that even after applying the luminosity criterion the range in colour for edge-on galaxies is still much larger than for face-ons. This clearly shows that at a given inclination the amount of extinction varies from galaxy to galaxy. The limit at $M_K = -22$ is not critical. The relations do not change very much after applying a different $M_K$ selection.

We conclude from Fig. 1 that galaxies seem to behave as expected if they are opaque in the centre and at $r_{eff}$ but are transparent at $D_{K,21}$.

## 3. Analysis of the scalelength ratios

In this section we use the slopes of the radial surface brightness profiles, or the scale lengths, to determine the amount and the distribution of the dust. We consider a galaxy that is representative of the sample as a whole: the ratio of exponential scale lengths in $B$ and $K$ $\alpha_B/\alpha_K$ equals 1.3, when face-on, and 1.7 when seen edge-on (Fig. 4 of Paper I).

The fits to obtain these scale lengths have been made on the radially averaged surface brightness profiles (Paper I), between 1 and 3 $K$-band scale lengths. Here, exactly the same radial range has been used for the fits in $B$ and $K$. First we have to correct for the effect of radial abundance gradients. Two independent lines of reasoning are used here (see Paper I). First, for galaxies similar to the ones discussed here, but without much dust, i.e. S0 or Sa galaxies, Balcells & Peletier (1994) find that the ratio of the scale lengths in $B$ and $I$ equals $1.04 \pm 0.05$. Using almost any stellar population model this gives a scale length ratio between $B$ and $K$ of $1.08 \pm 0.10$. The second argument comes from abundance gradients measured from HII regions. Literature data (Vila-Costas & Edmunds 1992; Zaritsky et al. 1994) indicate that the average gradient in normal nearby galaxies equals $\Delta(O/H) = -0.25$ dex/($B$ scale length), or $-0.17$ dex/($K$ scale length). This corresponds to a scale length ratio $\alpha_B/\alpha_K = 1.17$, using a simple single-age stellar population model. This corresponds more or less to the lower limit of our observations. We assume in the rest of this *letter* that stellar population gradients are responsible for a scale length ratio $\alpha_B/\alpha_K = 1.1$, independent of axis ratio. So the dust will only have to explain $\alpha_B/\alpha_K = 1.2$ for a face-on galaxy, and 1.6 for an edge-on. De Jong (1995) claims that the stellar population gradients in these spirals are much larger than assumed here. His reasoning is that scattering by dust is so important that the amount of extinction needed to explain the gradient in a blue colour like $B - V$ would imply a gradient in $B - K$ which is much larger than observed. Because of the discrepancy with metallicity measurements from HII regions he also needs considerable age gradients. His explanation may be valid for his data, consisting only of face-on galaxies, but with his inferred stellar population gradients one would not expect the good correlation with inclination that we see here. Given also the fact that there would be a discrepancy with S0's and Sa's (see above) we suggest that the explanation given here is more realistic.

To model the observed scale length ratios we first tried to use the relation of Valentijn (1991) between $\alpha_{obs}$, $\alpha_*$ and $\alpha_d$, the observed scale length, the stellar scale length, and the dust scale length. The model gives an analytic solution to a radial version of the equation of radiative transfer. It includes uniform distributions of dust and stars in separate layers. We found that the results depend quite strongly on the vertical distribution of the dust and stars, and since there is a large difference between uniform and exponential vertical distributions, we resorted to numerical modeling. Our exponential dusty galaxy model contains a uniform distribution of stars, exponential in the radial and vertical direction, and a uniform distribution of dust, also exponential in both directions, but with different scale lengths and heights. The ratio of the scale heights of dust and stars we call $\zeta$. We project the models and fit exponentials to the major axis profiles in the same way as is done for the observations. One set of models has been made for face-on galaxies and one for galaxies at 80 degrees.

The scale length ratios are plotted in Fig. 2 for various values of $\zeta$ and $\alpha_d/\alpha_K$. We impose (driven by the observations) the condition that $\alpha_B/\alpha_K$ in the face-on case lies between 1.1 and 1.3, and in the edge-on case between 1.5 and 1.7. Furthermore $\tau^{edge}$ should be 5 times as large as $\tau^{face}$, reflecting the difference in path length through the dust. Table 1 indicates which model galaxies fulfill the first criterion (column 3), which the second (column 4), and which ones fulfill the latter as well, i.e. are acceptable solutions (column 5).

**Table 1.** $\tau_{B,c}$ for those model galaxies which fulfill the first criterion ($1.1 < (\alpha_B/\alpha_K)^{face} < 1.3$; column 3), the second ($1.5 < (\alpha_B/\alpha_K)^{edge} < 1.7$ ; col. 4), and all three criteria, including the last one ($\tau^{edge} = 5\,\tau^{face}$), i.e. are acceptable solutions (column 5)

| Models: | | $\tau_{B,c}$ (face) | | | $\tau_{B,c}$ (edge) | | | $\tau_{B,c}$ (face) | | |
|---|---|---|---|---|---|---|---|---|---|---|
| $\zeta$ | $\alpha_d/\alpha_*$ | | | | | | | | | |
| (1) | (2) | (3) | | | (4) | | | (5) | | |
| 1 | 1 | 0.7 | – | 1.8 | 2.4 | – | 3.0 | | – | |
| 1 | 2 | 0.8 | – | 2.2 | 3.3 | – | 4.4 | 0.8 | – | 0.9 |
| 1 | 3 | 1.0 | – | 3.2 | 4.7 | – | 7.7 | 1.0 | – | 1.5 |
| 0.5 | 1 | 0.7 | – | 2.0 | 2.4 | – | 3.2 | | – | |
| 0.5 | 2 | 0.8 | – | 3.0 | 3.5 | – | 6.0 | 0.8 | – | 1.2 |
| 0.5 | 3 | 1.1 | – | $\infty$ | 6.2 | – | $\infty$ | 1.2 | – | $\infty$ |
| 0.3 | 1 | 0.8 | – | 2.3 | 2.4 | – | 3.2 | | – | |
| 0.3 | 2 | 0.8 | – | $\infty$ | 3.8 | – | 9.0 | 0.8 | – | 1.8 |
| 0.3 | 3 | 1.1 | – | $\infty$ | 9.0 | – | $\infty$ | 1.8 | – | $\infty$ |

We find solutions for any thickness, if $\alpha_d/\alpha_* \geq 2$. The central face-on optical depth is around 1 or 1.5. For a model with $\alpha_d/\alpha_K=2$, we would expect $\tau_B=0.2$-$0.3$ at 3 scale lengths for a face-on galaxy, or $\tau_B=1$ if the galaxy were edge-on. It is interesting to see that no solutions are found with $\alpha_d/\alpha_* = 1$.

The current model is too simplistic, especially because it ignores scattering, and because of the fact that dust is generally distributed non-uniformly across the galaxy.

The effects of scattering have been modeled by e.g. Witt et al. (1992) and Byun et al. (1994). The paper by Witt et al. shows that although it is difficult to calculate the effects of scattering one can describe it to a good approximation by modifying the extinction law, i.e. by decreasing the absorption in the optical relative to the IR (see also Rix & Rieke 1993).

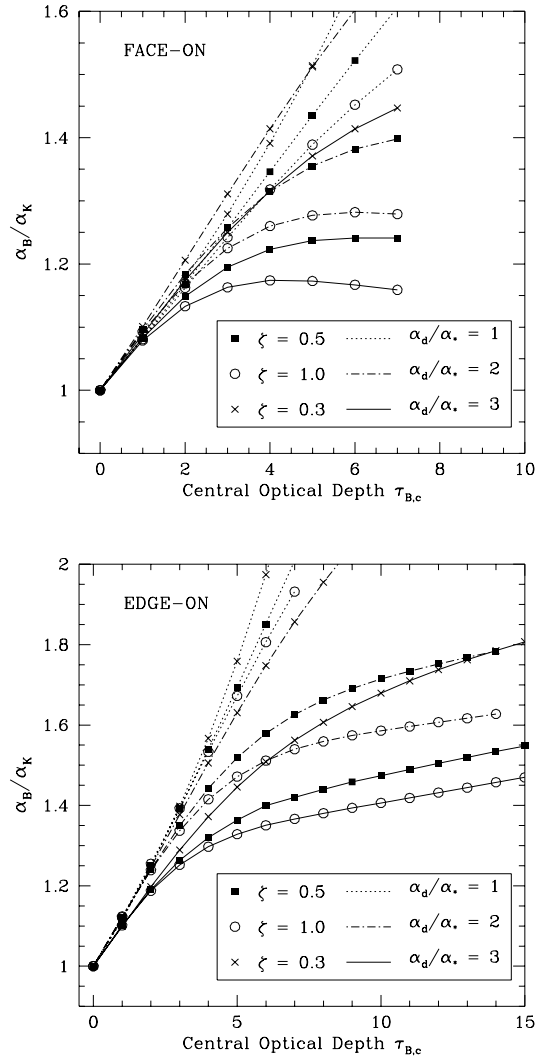

**Fig. 2.** Scale length ratios between $B$ and $K$ as a function of central optical depth for disk models with various ratios of scale lengths and heights of dust and stars.

Scattering is more efficient for face-on galaxies than for edge-on's since, in the first case, there is a larger probability that photons are scattered out of the plane towards the observer. Changing the extinction law by means of including the effects of scattering implies that more dust is needed to explain the behaviour in $B$ (a factor 2 in optical depth at most). The fact that scattering is more efficient for face-ons also means that a model including scattering requires less extinction for edge-on galaxies relative to face-ons than a model without scattering.

Let us now assume that the dust is distributed in clumps, as opposed to being uniform across the disk. We are interested in the case in which a substantial fraction of the disk is covered with dust clouds. If the fraction is much less, scale lengths are not affected much, contradictory to the current observations. A galaxy with a face-on cloud coverage of $\sim 50\%$ has an edge-on coverage of close to 100%. This means that the extinction is much more efficient for the edge-on than for the face-on since, in the latter case, never more than 50% of the light is being extincted. This means that the edge-on optical depth needed to produce the observed scale lengths is less than 5 times the face-on optical depth. How much less depends on the geometry of the dust.

When both effects are included, the models will need more face-on extinction, and a smaller ratio for face-on to edge-on. Taking as an example the solution for which $\tau^{edge}/\tau^{face} = 2.5$, and multiplying the face-on optical depths of Table 1 by a factor 2, the model is in agreement with the observations for a central face-on optical depth in $B$ between 2 and 3.

## 4. Discussion

In the previous section we have found that the only solutions that agree with the observed scale length ratios in $B$ and $K$ are solutions with large ratios between the scale *lengths* of dust and stars. Including absorption alone we do not manage to find a solution with equal scale length of dust and stars.

However, including scattering and clumpiness in the analysis, it is possible that one will find acceptable solutions with $\alpha_d/\alpha_* = 1$. These will have a central optical depth in B of about 2. In the outer parts, at 3 stellar scale lengths, the face-on optical depth would be reduced to 0.1.

What is the maximum amount of extinction at $D_{K,21}$? Here the scale length model of section 3 is less useful than the direct observations of section 2. Fig. 1 shows that models with $\tau_{21}^{face} \geq 0.5$ have big problems fitting the colours, especially when turned edge-on. Since the data here do not exclude $\tau_{21}^{face}$ between 0.1 and 0.5 it might be attractive to make a two-component model of the dust (see Valentijn 1995). A one-component model with a large ratio for $\alpha_d/\alpha_*$ (e.g. 3) would either have too little extinction in the centre or too much in the outer parts. Parameters of an acceptable two-component model are given in Table 2.

**Table 2.** Maximum amount of extinction for acceptable two-component models

| | $\tau_B^{face}$ at | | |
|---|---|---|---|
| $\alpha_{dust}/\alpha_\star$ | centre | $r_{eff}$ | $r_{B,25}$ |
| 1 | 2.0 | 0.7 | 0.1 |
| 3 | 1.0 | 0.8 | 0.4 |

An advantage of having these two-component models is that diameters like $D_{B,26}$ will have a smaller dependence on axis ratio (V94, Chołoniewski 1991). In the same way scale lengths, as derived from IRAS-CPC images, can be better understood (Van Driel 1995). This way, a randomly oriented galaxy would have $\tau_{B,21} \sim 1$.

How does this result compare to the current literature? In our approach, we compare light which suffers little from extinction (K-band) with light which is strongly affected by extinction (B-band) within individual galaxies. This is a more direct way of deriving the extinction as compared to a statistical approach, in which only light affected by extinction is investigated and the extinction is derived from studing galaxies of different inclinations. We find that at least the galaxies with inclinations $< 45°$ are optically thin at $D_{K,21}$ or $D_{B,25}$. But more inclined galaxies could well be opaque at that position, so that there is no conflict with Valentijn (1994) or Bottinelli et al. (1995), who for large samples in $B$ show that diameters at $\mu_B = 25$ mag (arcsec)$^{-2}$ do not vary with inclination. Our results are in agreement with Giovanelli et al. (1994) who show that at two or three scale lengths spiral galaxies are transparent in $I$. Using the galactic extinction law (see Knapen et al. 1991; Jansen et al. 1994) $\tau_B/\tau_K \sim 2.2$, implying that for our sample $\tau_I$ at $D_{K,21}$ is at most 0.2, in agreement with Giovanelli et al. . They also show that the scale lengths increase with axis ratio, more or less in agreement with the present paper.

We find here that there is in general a considerable amount of extinction in spiral galaxies in the central regions, all the way out to $r_e$. In the outer parts of our data range, at $D_{K,21}$, or at 3 stellar scale lengths, we find a maximum optical depth of 0.5 in $B$ for a face-on galaxy. Assuming that the dust is distributed radially in the same way as the stars, this upper limit would go down to 0.1. We find that these galaxies in general, except for some edge-ons, are not optically thick at $D_{K,21}$, although some dust could still be present there.